\pdfoutput=1
\documentclass{JINST}

\newcommand{\bdec }{\mbox{$\beta$~decay}}

\newcommand{\ttwo }{\mbox{$\rm T_2$}}
\title{A mobile Magnetic Sensor Unit for the KATRIN Main Spectrometer }

\author{{A. Osipowicz$^a$\thanks{Corresponding
author.}, W. Seller$^a$, J. Letnev$^a$, P. Marte$^a$, A. Müller$^a$, A. Spengler$^a$, A. Unru$^a$} \\
\llap{$^a$}University of Applied Sciences, Fulda, Germany \\
  E-mail: \email{Alexander.Osipowicz@et.hs-fulda.de}}

\abstract{The KArlsruhe TRItium Neutrino experiment (KATRIN) aims to measure the electron neutrino mass with an unprecedented
sensitivity of 0.2 eV/c$^2$, using \bdec~electrons from tritium decay. For the control of magnetic field in the  main spectrometer area of the KATRIN experiment a mobile magnetic sensor unit is constructed and tested at the KATRIN main  spectrometer site. The unit moves on  inner rails of the  support structures of the low field shaping coils which are arranged  along the the main spectrometer. The unit propagates on a caterpillar drive and contains an electro motor, battery pack, board electronics, 2 triaxial flux gate sensors and 2 inclination senors. During operation all relevant data are stored on board and transmitted to the master station after the docking station is reached.}


\begin{document}

\section{The KATRIN setup}

\begin{figure}[h]
\centering
\includegraphics[width=.8\textwidth]{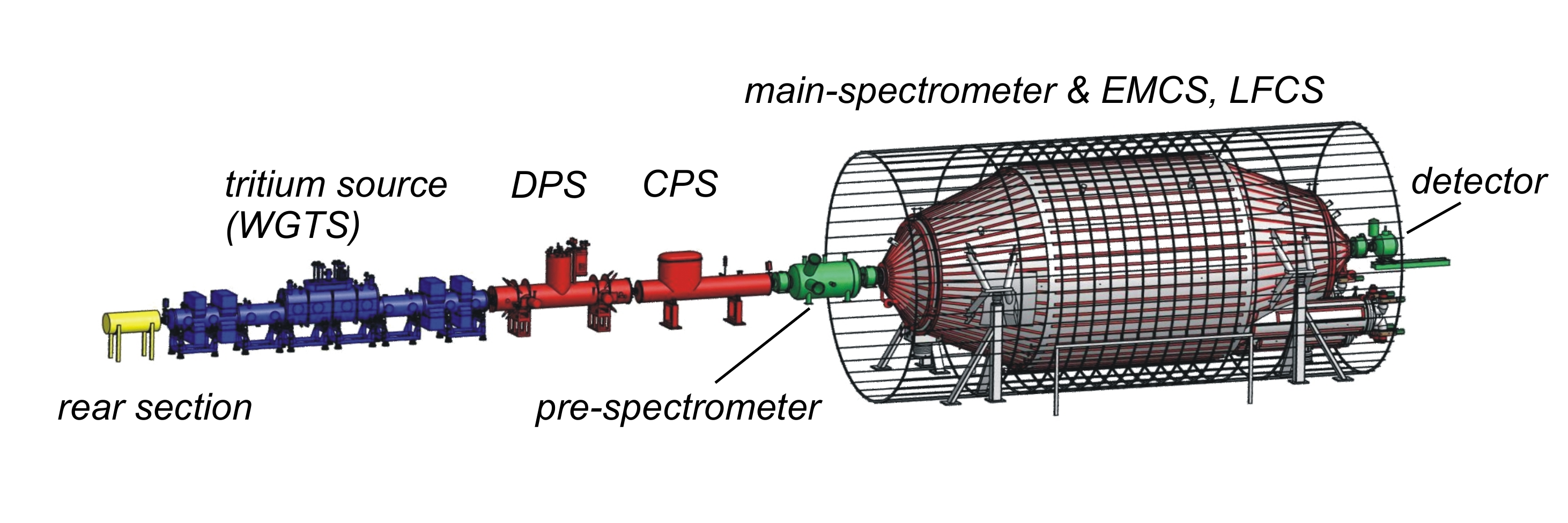}
\caption{\label{fig:KATRIN-all-eng}Schematic view of the KATRIN experiment (total length 70 m) consisting of calibration and monitor rear system, 
with the windowless gaseous \ttwo -source (WGTS), differential pumping (DPS) and cryo-trapping section (SPS), the small pre-spectrometer and the large main spectrometer with the large magnetic coil systems to compensate the earth magentic field (EMCS) and to shape the magnetic transport flux (LFCS) and lastly the segmented PIN-diode detector.}
\end{figure}

The KArlsruhe TRItium Neutrino  experiment \cite{katrin-loi} (see Fig.\ref{fig:KATRIN-all-eng})  is set up at the Karlsruher Institute of Technology (KIT), Germany. It is designed to measure the mass of the electron neutrino in a direct and model-independent way with a
sensitivity of $m_{\nu}=0.2$ eV/c${^2}$ (90\% confidence level) from tritium \bdec \cite{katrin-loi}. KATRIN uses a magnetic transport field  that connects the source and detector in combination with  integrating electrostatic energy filters (MAC-E-spectrometers). 
Conceptual essentials of the MAC-E spectrometer\cite{Pic92,Lob85} are the  magnetic field gradients in pre - and main-spectrometer that adiabatically convert cyclotron energy $E_{cyc}$ into energy $E_p$ parallel to the magnetic field lines and vice versa. 

\begin{figure}
\centering
\includegraphics[width=0.9\textwidth]{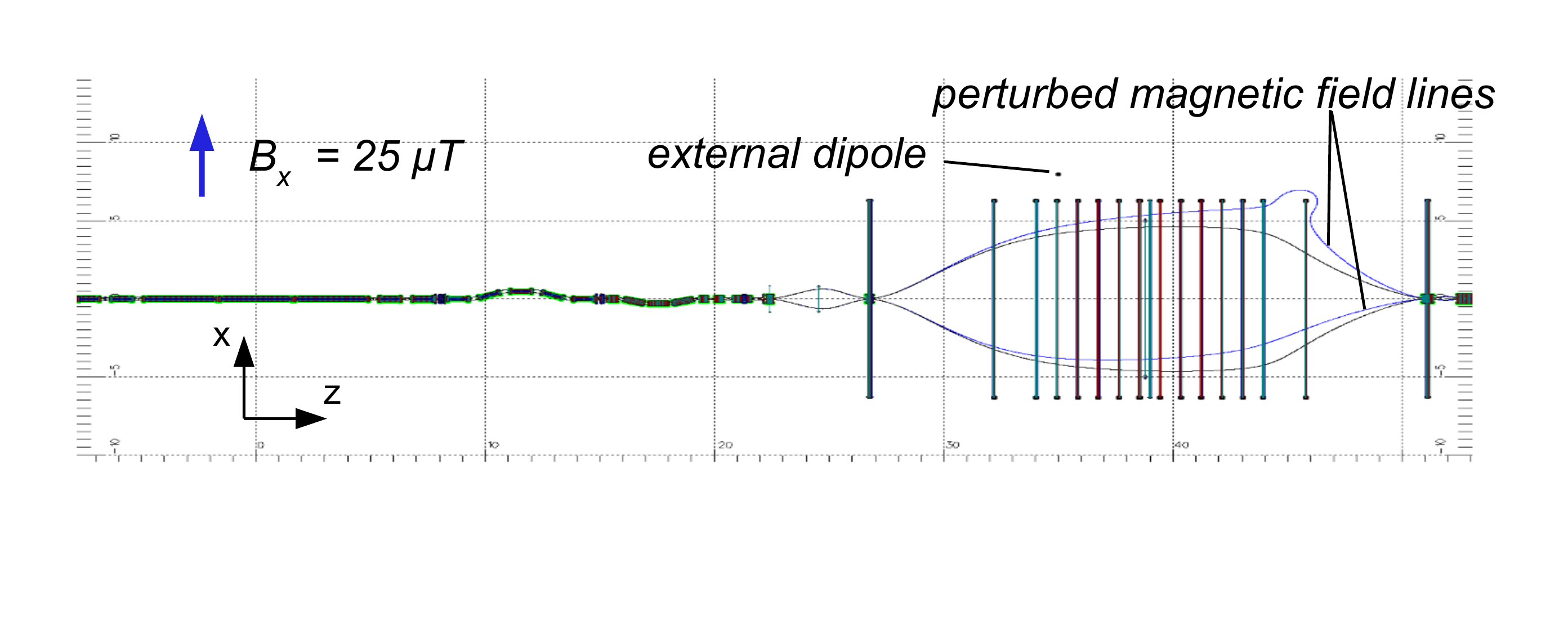}
\caption{\label{fig:B-all-off-on} A CAD bird view (taken from a PartOpt \cite{PO} simulation) of the energized
KATRIN solenoid chain and LFCS. The analyzing area is situated at the center of the main spectrometer  where the shape of the magnetic  flux is mainly given by the spectrometer solenoids and the LFCS. In this simulation the extreme magnetic flux lines ($191$ T cm$^2$) have been tracked in the horizontal plane with and without the perturbing influence of the earth magnetic field  $B_x = 25$ $\mu$T  and a small external dipole with $600 $ $\mu$T central induction. The perturbed 
flux lines are indicated.}
\end{figure}

In the minimal magnetic field (the analyzing field $B_A\approx 3-6$ $\mu$T and $d_\Phi=4.5$ m) at the center of the MS, a retarding electric field distribution allows  an integral energy analysis of $E_p$. The shape of the magnetic flux tube in the MS area defines the magnetic resolution, i.e. the amount of residual cyclotron energy $E_{cyc}$ that can not be analyzed and thus strongly influences  the resolution function. Systematic error considerations \cite{Tit} demand a  homogeneity of the magnetic field distribution in the analyzing volume of $\Delta(B)/B<0.04$.   Moreover, the  alignment of magnetic field lines plays a crucial role in the production of secondary electrons and electronic background either through traps or wall contact (see Fig. \ref{fig:B-all-off-on}).

Large coil systems \cite{OSI} are arranged around the MS for a) global magnetic field compensation, e.g. earth magnetic field (EMCS) and b) fine tuning of the magnetic transport flux with a set of large circular low field coils (LFCS) mounted coaxially with the MS.
However, possible influences of residual external dipoles, magnetization  in the MS environment by  the  high field solenoids and/or EMCS, LFCS and the correct orientation of the  spectrometer solenoids have to be controlled. Due to the extreme MS vacuum conditions  the installation of magnetic sensors inside the MS  is not possible.\par

\begin{figure}[h]
\centering
\includegraphics[width=0.7\textwidth]{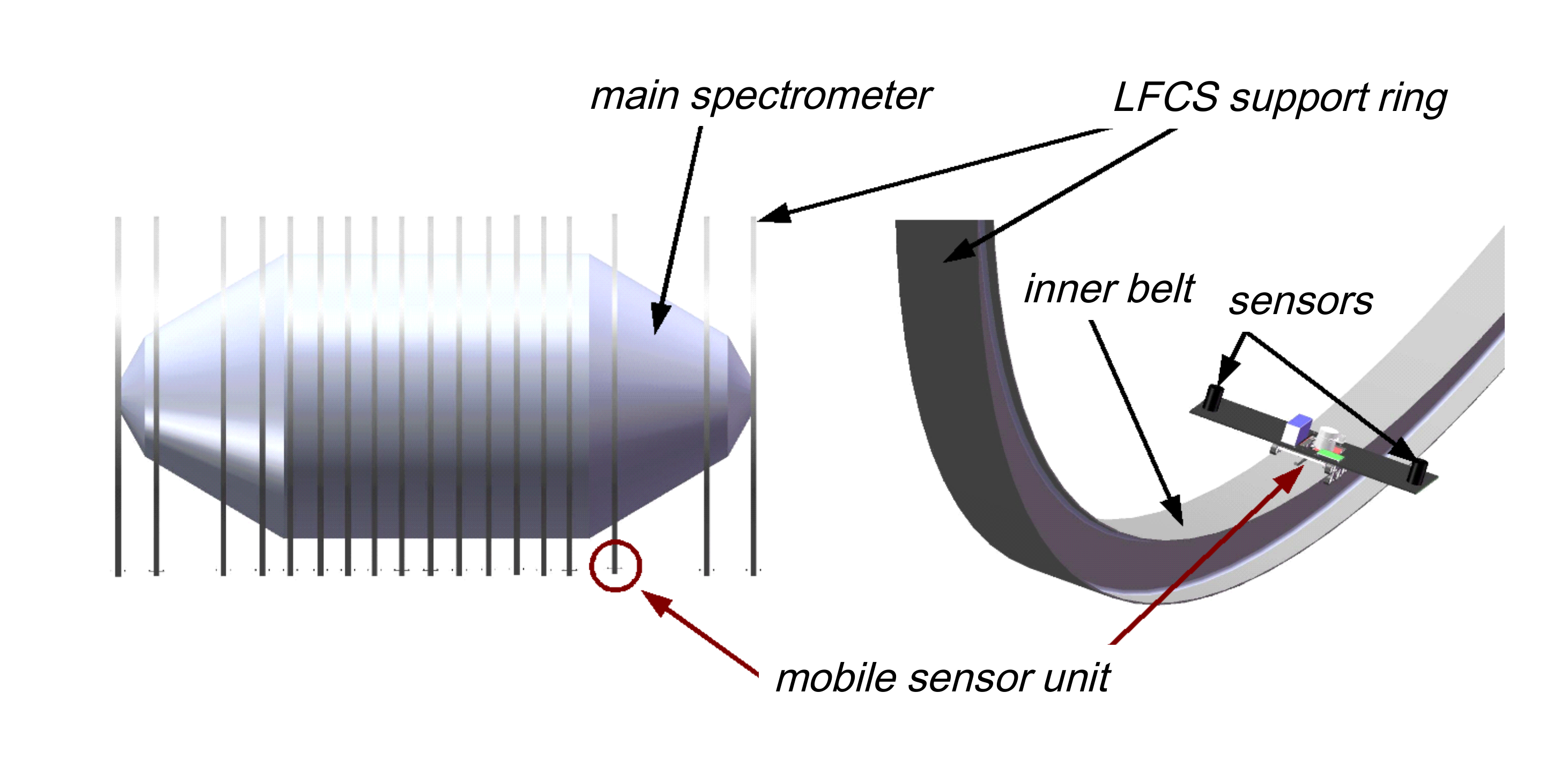}
\caption{\label{fig:MOBS-support-ring}View of the main spectrometer tank with the LFCS ring system. Right: The mobile sensor unit with 2 sensors on the inner belt of a LFCS support ring. }
\end{figure}

In this paper we present a mobile sensor unit, designed to  move along the inner belts of the LFCS support structure (see Fig.\ref{fig:MOBS-support-ring}), close to the outer MS surface, but well inside the EMCS, LFCS current lines. It allows to sample a large number of magnetic field values from large areas of MS surface.

\section{\label{sec:2}The Mobile Sensor Unit}

\subsubsection{Mechanical structure}

\begin{figure}[h]
\begin{minipage}{0.48\textwidth}
\centering
\includegraphics[width=1.\textwidth]{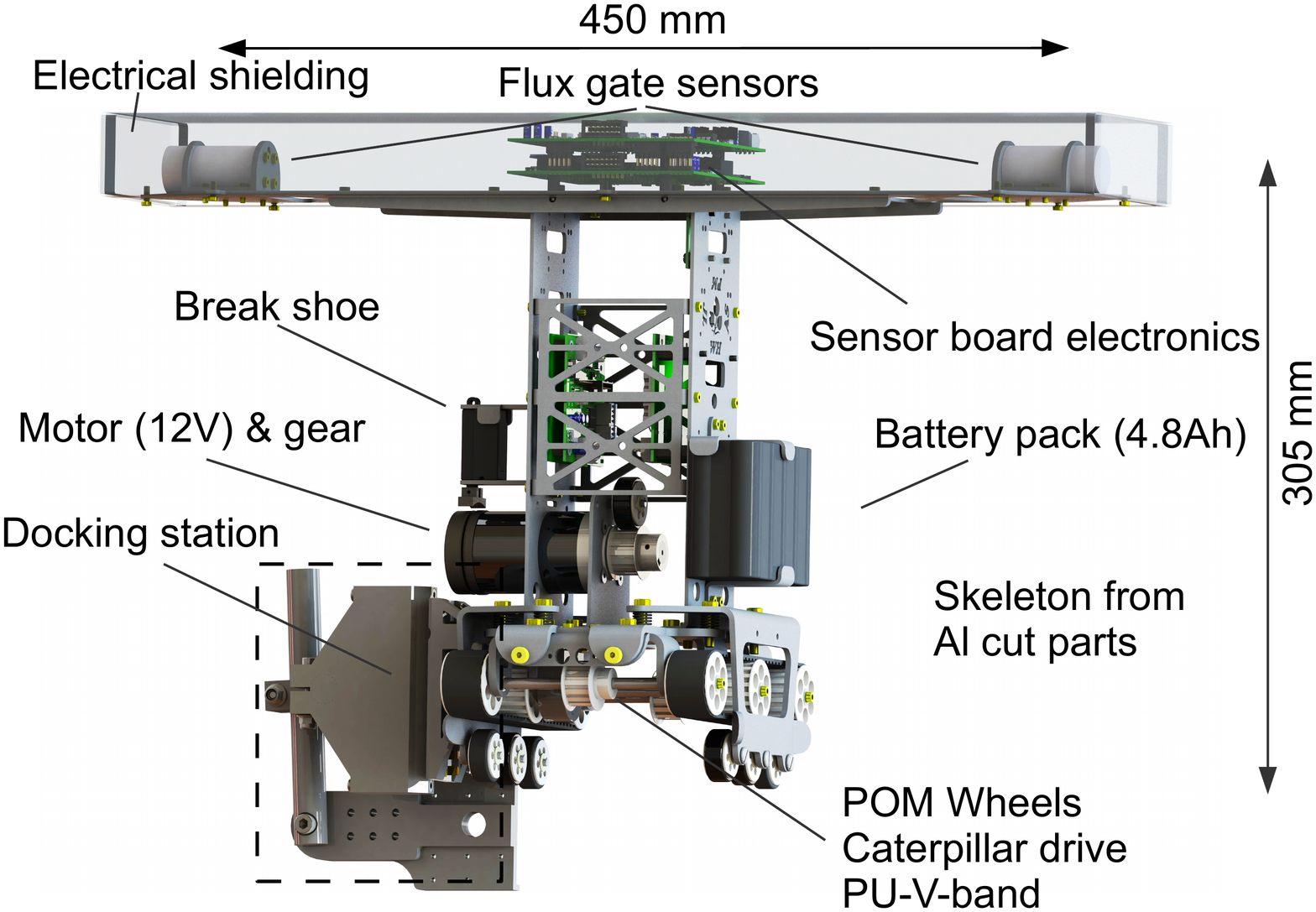}
\caption{\label{fig:MobS-cad} CAD view of the MobS with docking station. The chassis suspension with a caterpillar drive, the tower with motor, gear and battery pack. The  electrically shielded  control  board on top contains board electronics and two triaxial flux gate and 1 dual inclination sensor. The electrical shield is sketched transparently.}

\end{minipage}\hfill
\begin{minipage}{0.48\textwidth}
\centering
\includegraphics[width=1.\textwidth]{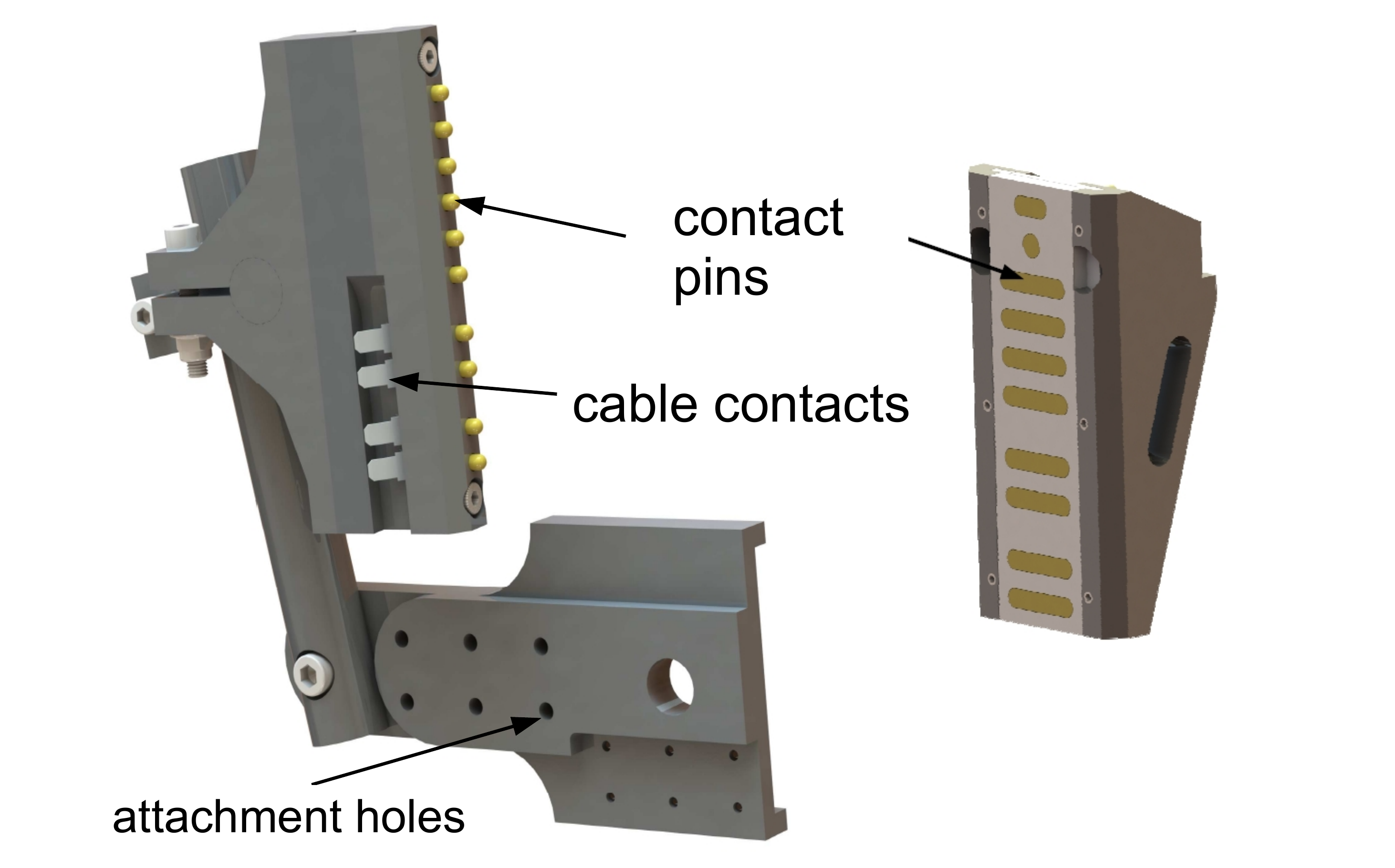}
\caption{\label{fig:DS-wing-all-eng} CAD view of the docking station (DS) (left) and docking block (right). The block on is attached at the MobSU. The contacts from top: 2 detection contact pads, 2 RS 485 data lines, 2 RS 232 data lines, 2 power supply  and 2 charging lines.}

\label{fig:startcoords}
\end{minipage}
\end{figure}

To minimize self-magnetization,  the skeleton of the mobile sensor unit \cite{UNR, LET} (MobS)(see Fig. \ref{fig:MobS-cad}) is assembled from aluminum cut parts and  consists of a chassis  with a drive, a tower and the wing-shaped sensor board that is attached on top of the tower.  The caterpillar drive  consisting  of 3 acetal resin wheels  (and 3 counter wheels per side)  bearing a polyurethane toothed belt is chosen to ensure slip free motion on interfaces between the LFCS ring arcs . The tower hosts a brush-less DC motor (maxon motor EC45, 12 V/30 W),  gear (maxon motor gear GP 42 C), and rechargeable battery pack (Lead AGM, 1.8 Ah). To avoid currents during the measurement intervals  a linear actuator is used  as a  break shoe that mechanically clamps the motor. 


The sensor board attached at the top of the tower is equipped  with two triaxial custom designed flux gate sensors  (see Table \ref{tab:sensors}), a dual axis digital inclinometer (Analog Device ADIS 16209), and two electronic boards. The flux gate sensors are positioned along common axes within $10$ $\mu$m tolerance.   As the sensor board is facing inward to the main spectrometer surface, which can bear voltages up to 32 kV, the sensor board is electrically shielded. Mounted at one side of  the chassis suspension is the docking block that detects the docking station (DS) via  2 sliding contacts (see Fig. \ref{fig:DS-wing-all-eng}). The DS is permanently  attached to the LFCS inner belt. Both docking block and docking station are manufactured from PVC with contacts for data, power and charging lines.\par


\subsubsection{Electrical layout}

\begin{figure}[h]
\centering
\includegraphics[width=0.6\textwidth]{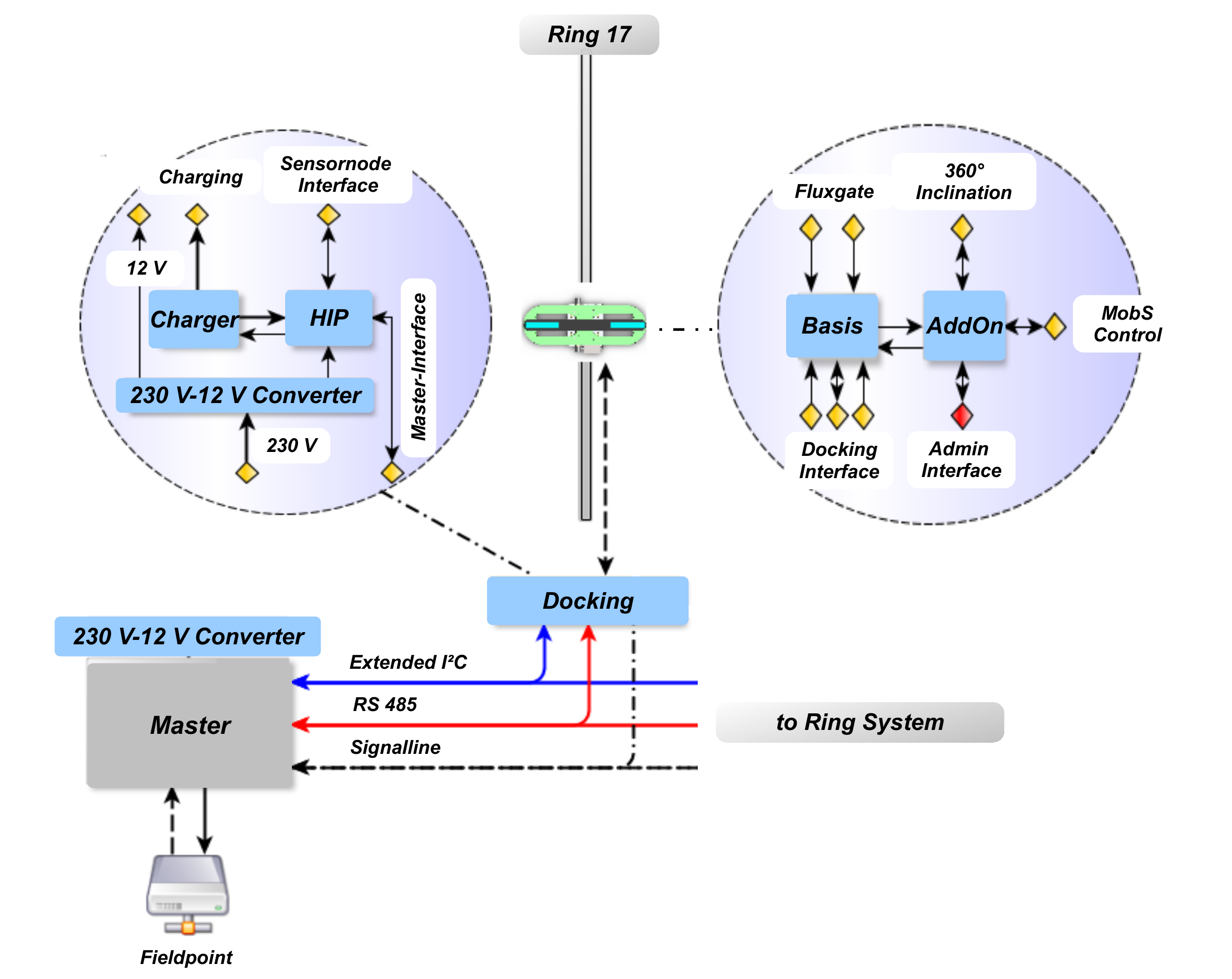}
\caption{\label{fig:system-one} Sketch of the electrical layout of sensor board electronics (right circle), docking station electronics (left circle), data and power lines to master module and Fieldpoint to KATRIN slow control (below). The layout is designed to incorporate up to 20 MobS on further LFCS rings.}
\end{figure}

The basis board (see Fig.\ref{fig:system-one}) containing  a microcontoller unit (TI MSP 430F5438) reads the flux gate signal via precision operational amplifiers (LINEAR TECHNOLOGY LT6013/LT6014) and analog to digital converters (TI TLC 3574, 14-bit) and controls the actions of the MobS. The AddOn board hosts 6 MB of flash memory, one inclination sensor,  USB interface, I$^2$C interface and the motion control module (MCM) (Atmel MCU Atmega 8). The MCM controls the drive chain and the linear actuator (break shoe). It also controls the incremental encoder that gives 2048 impulses per wheel  revolution (radius wheel $R= 32,75$ mm) and allows a positional resolution of $50.16 \: \mu\rm{m} \pm 25.8 \:\mu\rm{m}$.

The docking station is designed   a) to act as an interface in the  data channels  and b) as a charging station for the battery pack. This is realized with a dedicated circuit board (HIP) based on a Atmel MCU Atmega 328.
The master module (Master) serves as a data collector and provides an interface to the KATRIN database.       

\subsubsection{The sensor coordinate systems}

\begin{figure}[h]
\centering
\includegraphics[width=0.7\textwidth]{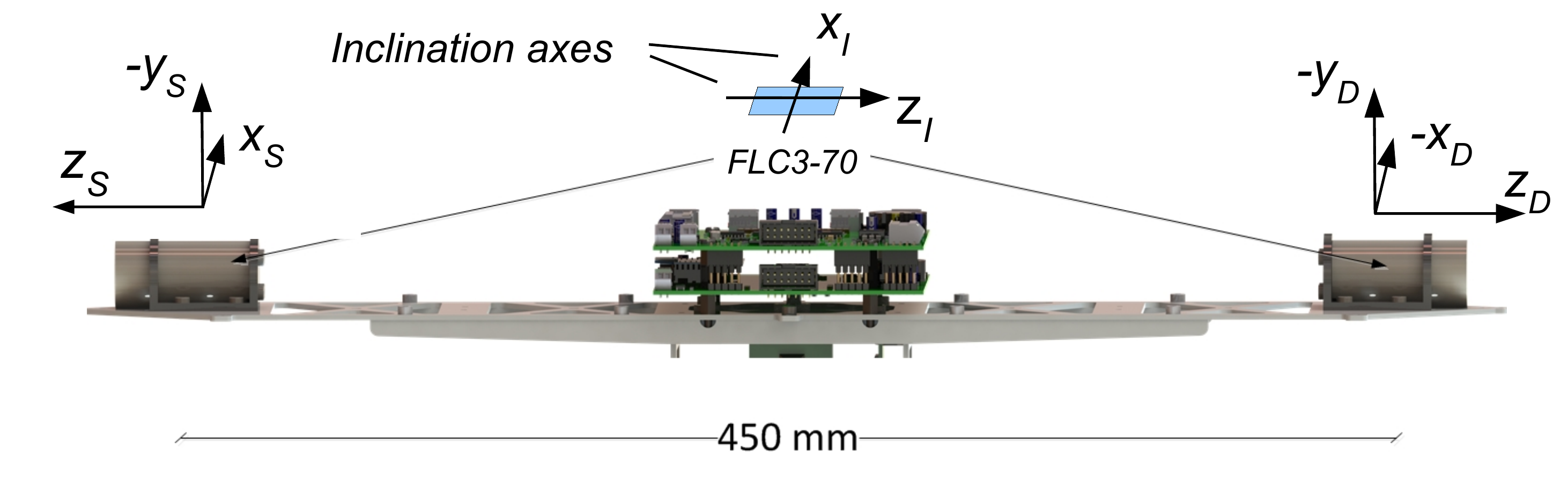}
\caption{\label{fig:Sensor-coords} The coordinate systems of the individual sensors that are aligned along a common $z$ -axes, which is parallel to the global $z_g$-axes.  The sensors are labeled according to their orientation with respect to the source (S) and detector (D).  The MobS is moving in $x_S$, $-x_D$ -direction. The inclination sensor axes $x_I$, $z_I$ are oriented parallel to the $x$, $z$ -axis respectively. }
\end{figure}

The orientation of the sensor coordinate systems on the MobS is shown in Fig.\ref{fig:Sensor-coords}.
As the MobS is moving along a LFCS ring, the sensor coordinate systems change their position and orientation with respect to the global KATRIN cylindric coordinate system $z_g, \varphi_g,r_g$. With  $z_g$, the central axis pointing from source to detector, $r_g$ the radial coordinate and $\varphi_g$ the azimuthal coordinate, the MobS will pick up  axial, azimuthal and radial magnetic field components. However, the exact anchoring  of the local coordinate systems into the global system has not been carried out yet as the MobS track is not ideally ring shaped and shows local misalignments.

\section{\label{sec:level1}The MobS residual magnetic field} 
In order to determine possible residual magnetic field components $\vec{B}_{res}$ that are produced by the MobS itself a, difference measurement in an iron free environment was performed. After positioning on a CNC machined Al-turntable, the MobS was oriented horizontally by inclinometer reading. Their tilt sensing system uses gravity as its only stimulus and is specified to have an angular accuracy of $0.1^{\circ}$.

\begin{figure}[h]
\begin{minipage}{0.48\textwidth}
\centering
\includegraphics[width=.85\textwidth]{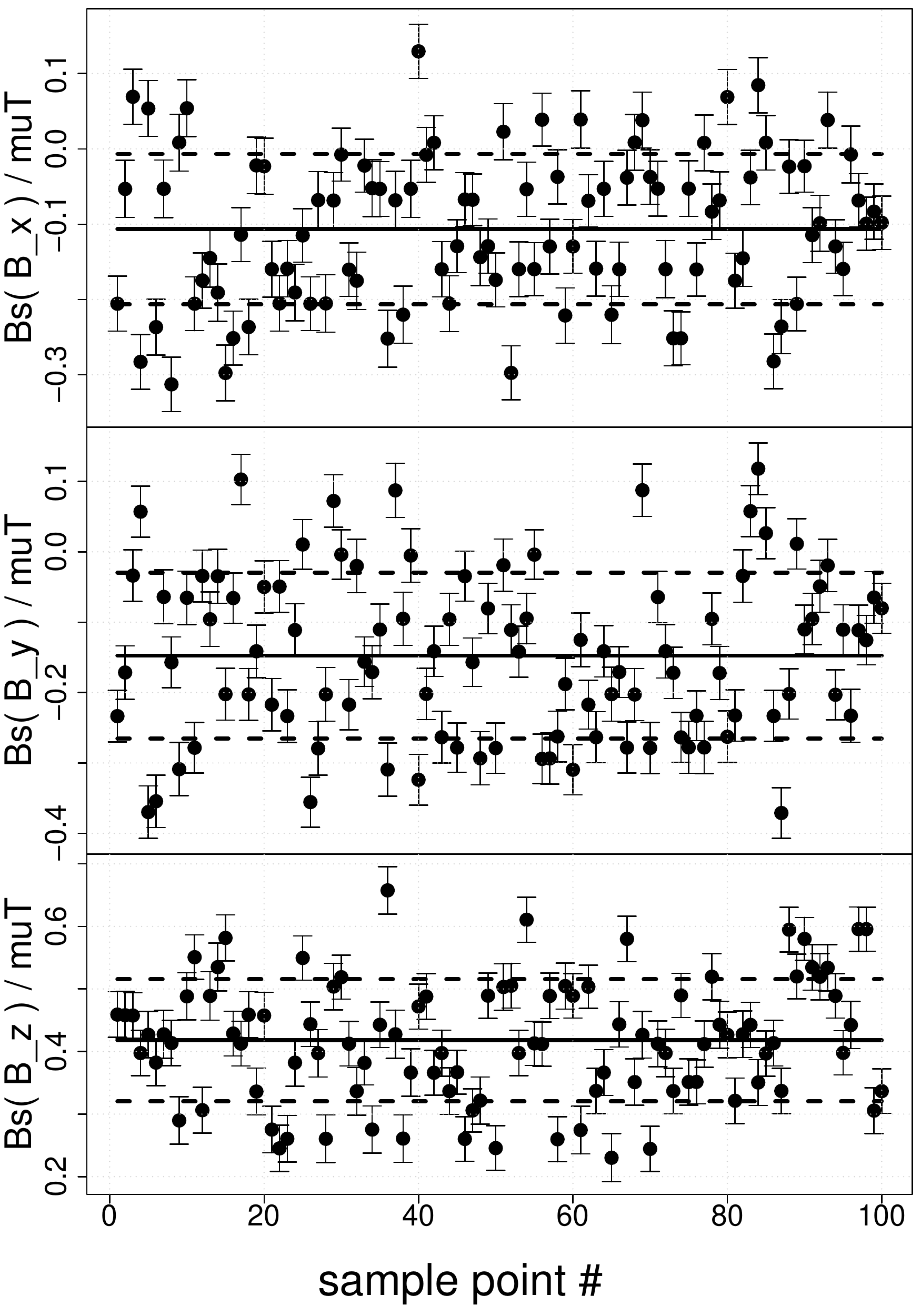}
\end{minipage}\hfill
\begin{minipage}{0.48\textwidth}
\centering
\includegraphics[width=.85\textwidth]{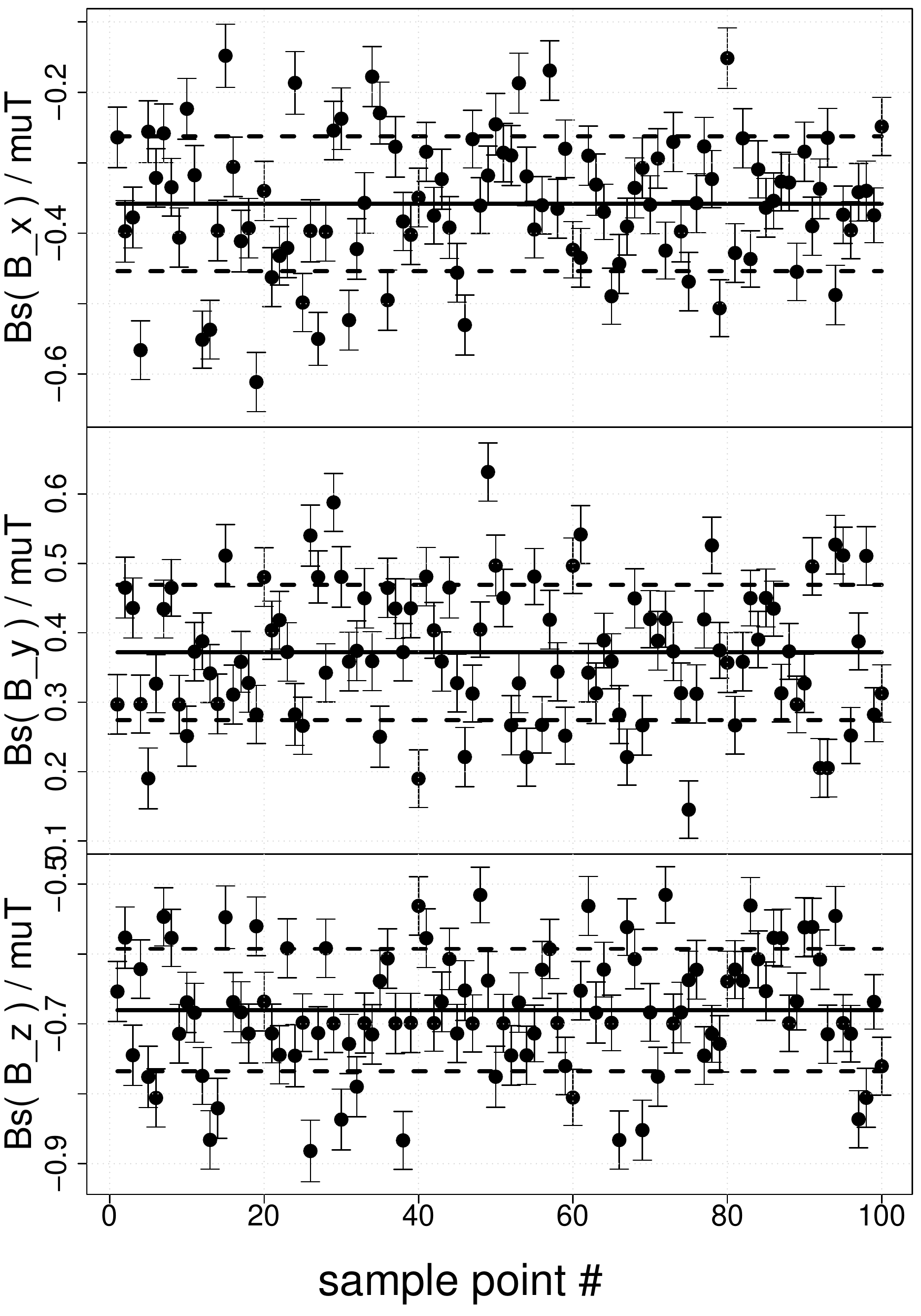}
\label{fig:startcoords}
\end{minipage}
\caption{\label{fig:Tor-SS} The  differences   between the magnetic field components measured in anti-parallel directions for  the source sensor (left) and detector sensor (right). The resulting mean and sigma is shown as a black line  and dashed line respectively.}
\end{figure}

With the FL3-500 sensor 100 magnetic  field samples are taken (conversion (DAC) rate $100 $ kHz, 64 conversions per sample point), then the table was turned by $180^{\circ}$ and the next 100 samples are taken. The resulting differences (see Fig.\ref{fig:Tor-SS})for the magnetic field components are  $\vec{B}_{res} = (-0.36\pm 0.1 \:\mu \rm{T},0.37\pm 0.1 \:\mu \rm{T},-0.68\pm 0.09 \:\mu \rm{T})$ for the detector sensor and $\vec{B}_{res}=(-0.11\pm0.1 \:\mu \rm{T},-0.15\pm0.12 \:\mu \rm{T},0.42\pm0.1 \:\mu \rm{T})$ for the source sensor.

\section{\label{sec:level1} Test measurement at the KATRIN site}

\begin{figure}[h]
\begin{minipage}{0.5\textwidth}
\centering
\includegraphics[width=.9\textwidth]{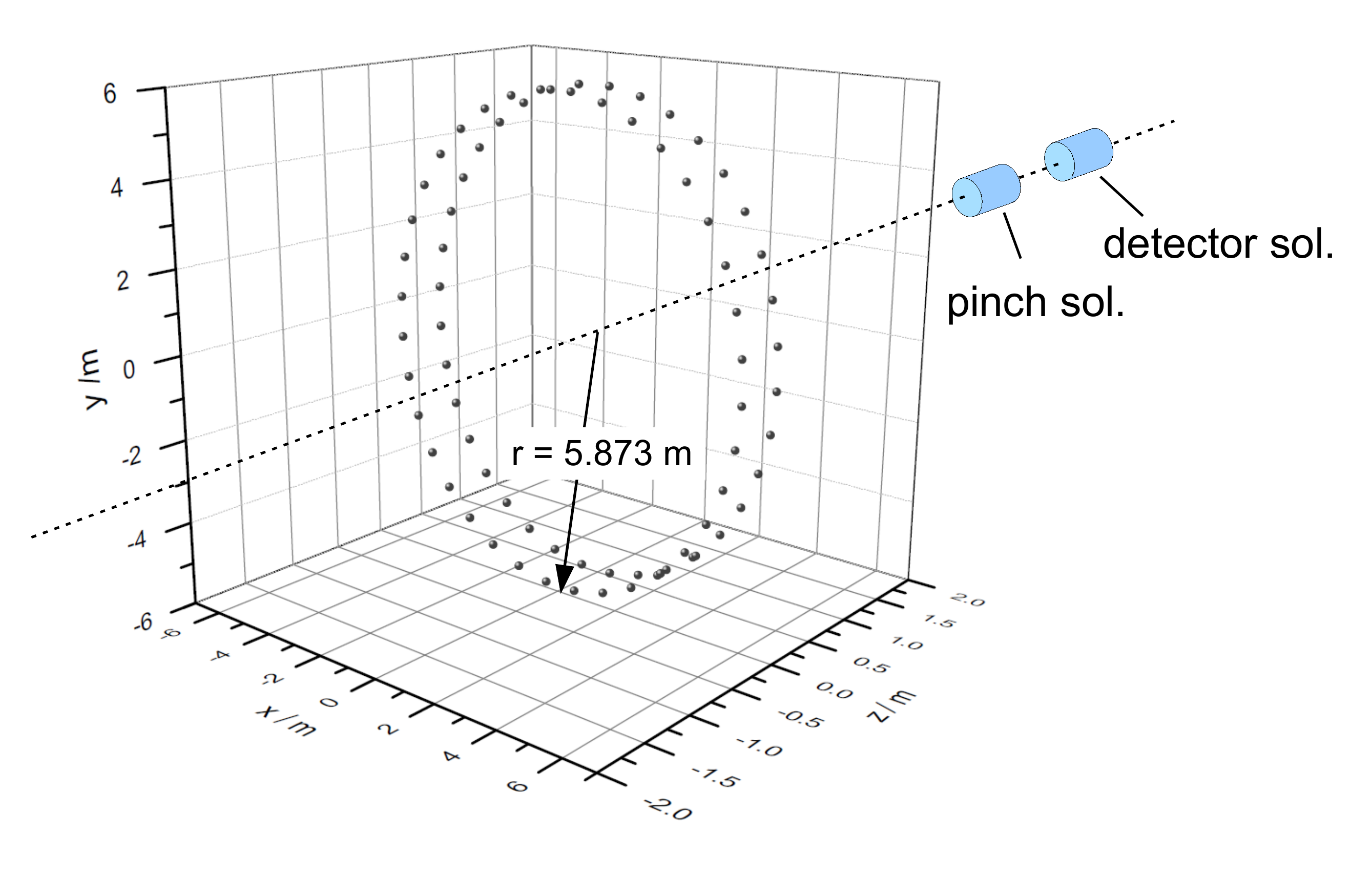}
\caption{\label{fig:sample-points-ext} The distribution of the 2*36 sample points along the track. The LFCS ring 17 is centered at the $z=0$ m position. Due to manufacturing tolerances the radial distance of the sensor positions of $r= 5.873$ m shows variations up to $0.05$ m along the arc.  The coaxial position of the pinch ($z_p = 12.4575$ m) and detector solenoid ($z_d=14.0575$ m) is indicated on the global z-axis (dashed line).}
\end{minipage}\hfill
\begin{minipage}{0.48\textwidth}
\centering
\includegraphics[width=0.8\textwidth]{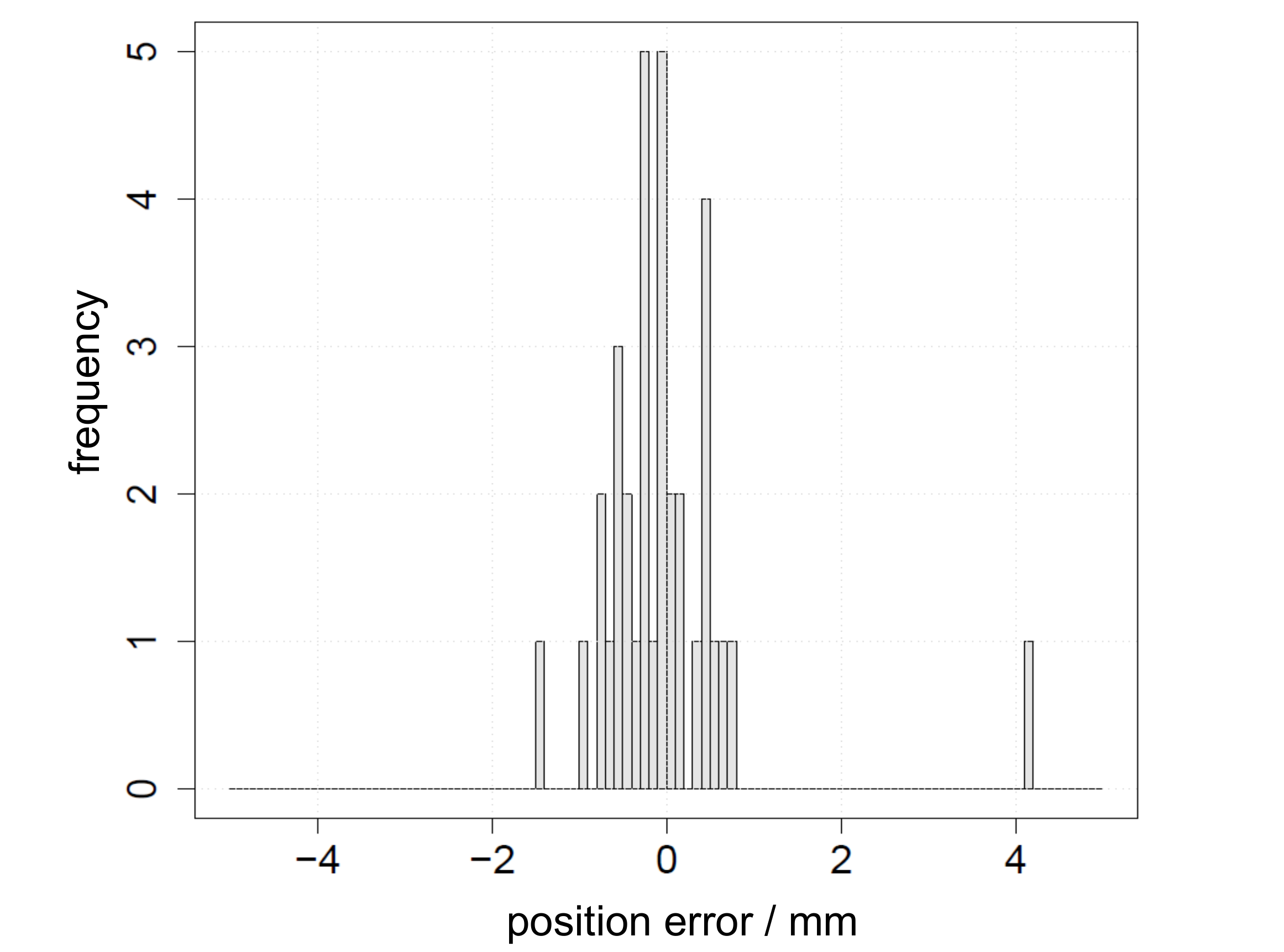}
\caption{\label{fig:Pos-Acc} The distance between two stops during one cycle  according to the step motor data shows a distribution that is caused by varying motor reaction times at different MobS position along the ring.}

\label{fig:startcoords}
\end{minipage}
\end{figure}

The magnetic field components are measured at 36 equidistant positions (see Fig.\ref{fig:sample-points-ext})   on the circumference of ring 17 (see Fig.\ref{fig:MOBS-support-ring}) at the KATRIN MS site. The positions are accurate within  $0.01$ m (see Fig.\ref{fig:Pos-Acc}).

\begin{figure}[h!]
\centering
\includegraphics[width=0.5\textwidth]{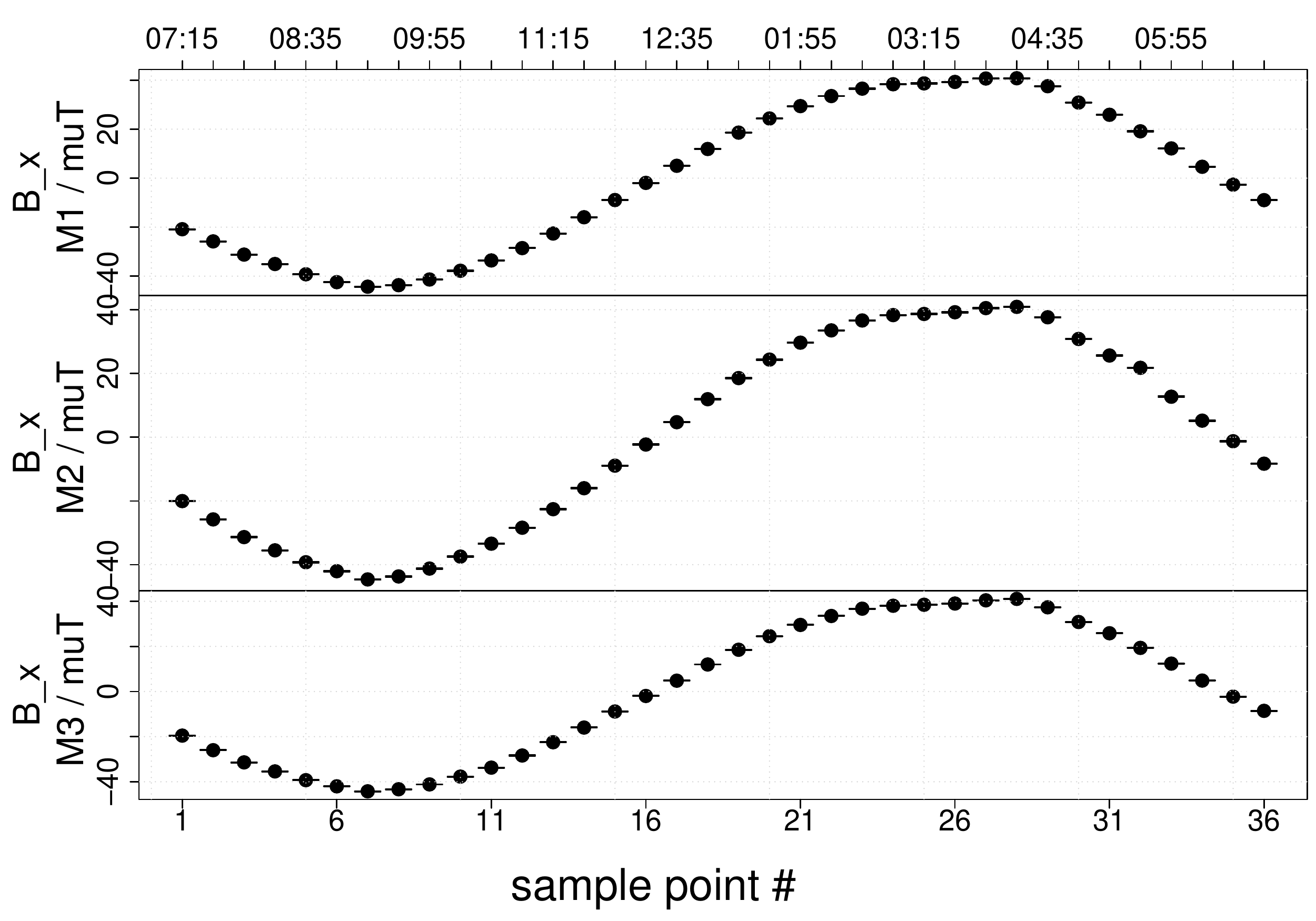}
\caption{\label{fig:Bx-0T-DS} The raw data for the $B_x$-component for the 3 runs with the magnet setting 0 T/0 T (background measurement) from the detector sensor. Each point represents the mean of 64 samples resulting in a small error. The position of the sample points on the arc is also given in "clock"-positions (at the top). In detector direction the Mobs performs a clockwise motion starting at the 7:15 position ending at the 6:55 position.}
\end{figure}

The samples are taken for 4 successive  pinch/detector (see Fig.\ref{fig:sample-points-ext}) solenoid current settings: a.)  central induction  $6$ T/  $3.6$ T, b.)  $3$ T/ $ 1.8$ T, c.)  $1.5 $ T/ $0.9$ T and  d.) $0$ T/ $0$ T allowing a background measurement. The two  solenoids are arranged coaxially with the LFCS coil with a central $z$-distance of $z_p = 12.4575$ m for the pinch and $z_d=14.0575$ m for the detector solenoid (distance values according to the detector installation position on 1st March 2012).    Fig.\ref{fig:Bx-0T-DS} shows the $B_x$-components for the three successive runs with the  $0$ T/ $0$ T magnet setting.  For each setting the MobS has completed 3 cycles. The time for completion of one cycle is about 5 minutes. 

\begin{figure}[h!]
\centering
\includegraphics[width=0.4\textwidth]{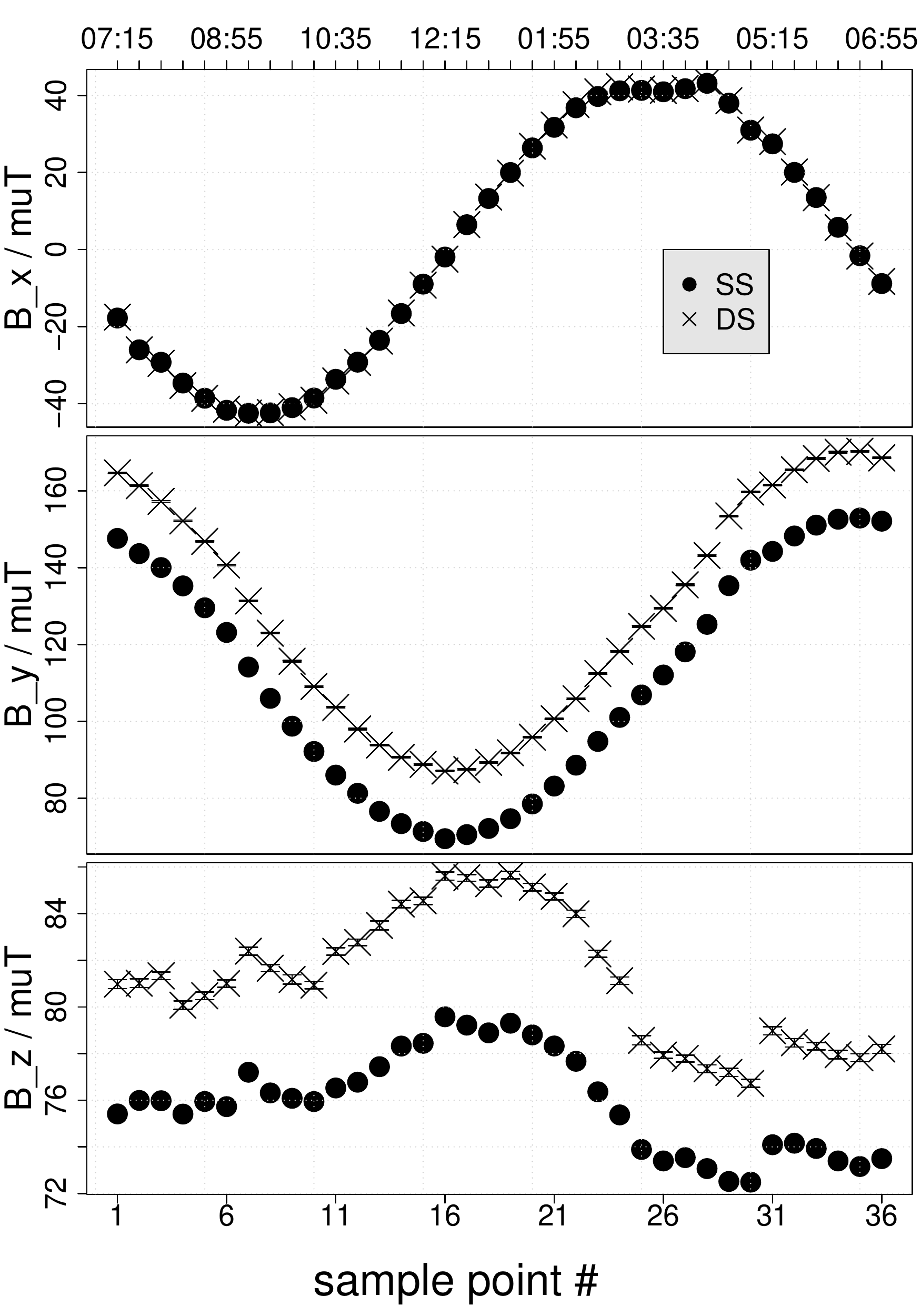}
\caption{\label{fig:Bxyz-6T-SS-DS} A compilation of the magnetic field components  from the detector sensor (blue crosses) and the source sensor (full dots) for the $6$T/$3.6$T magnet setting. As the detector sensor is closer to the detector it senses higher values for the axial $B_z$ and the radial $B_y$ component.  The  $B_x$ component readings are very similar, because the detector solenoids  do not produce any azimuthal field.}
\end{figure}

After statistics and correction for zero point and internal orthogonality error the values of $B_x$ and $B_z$ of the detector-sensor have been multiplied by $-1$ for display reasons (see Fig.\ref{fig:Bxyz-6T-SS-DS}).

\begin{figure}[h]
\centering
\includegraphics[width=0.5\textwidth]{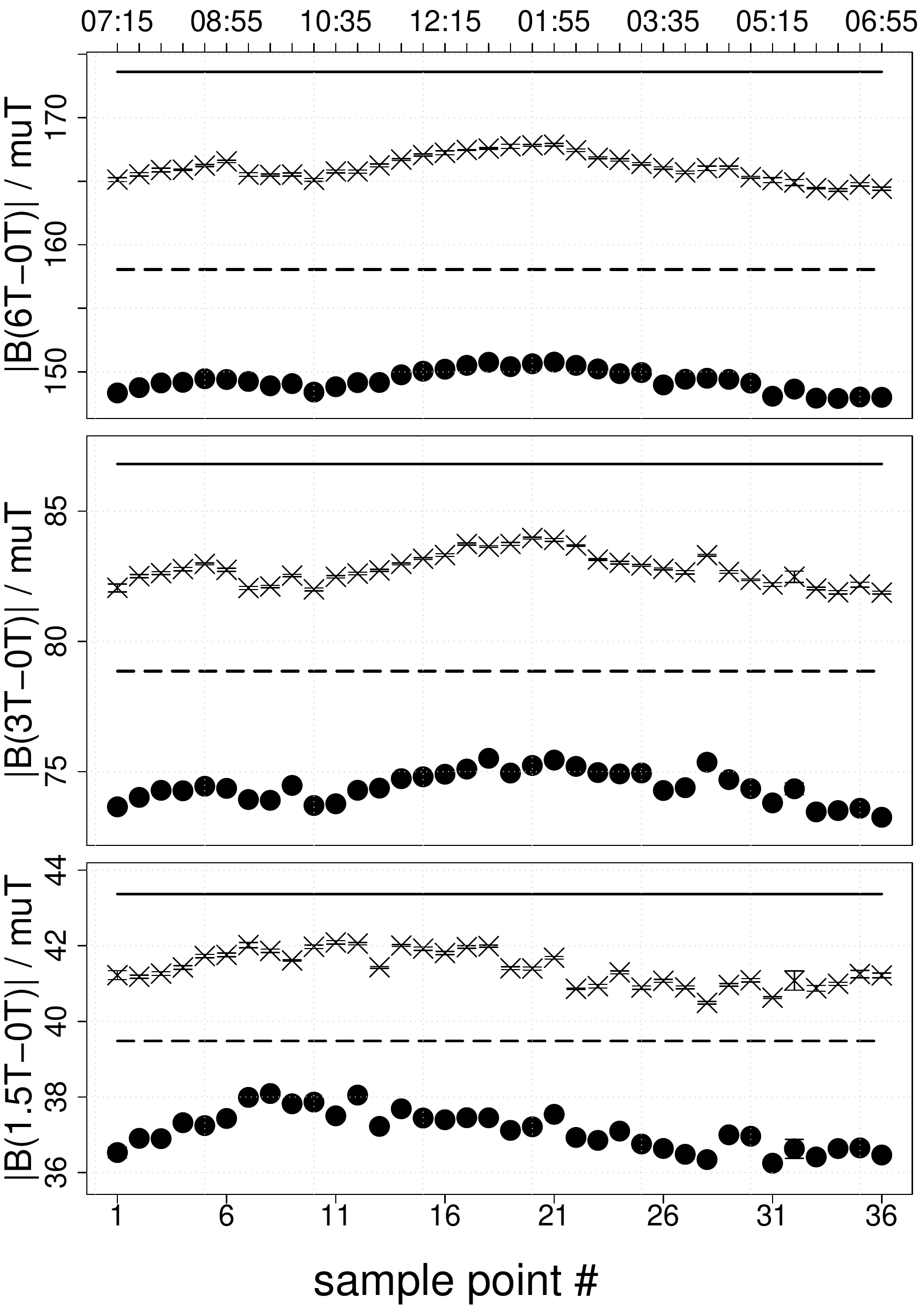}
\caption{\label{fig:B-6T-SS-DS} A comparison of the total magnetic field strength with simulated magnetic field values for the source sensor (full circles and dashed line) and detector sensor (crosses and  full line). Lower box: the $1.5$ T/ $0.9$ T, middle box: the $3$ T/ $1.6$ T and upper box; the $6$ T/ $3.6$ T magnet setting.}
\end{figure}

After subtracting the background field components from the data with energized detector solenoids, the resulting values can be compared to a magnetic field simulation (see Fig.\ref{fig:B-6T-SS-DS}). The geometric data of pinch and detector solenoid are taken from the manufacturers data sheet.     

The experimental data  show  a variation of up to $3 \;\mu$T  along the track for each detector magnet setting, moreover it is obvious that the difference between the simulated and the measured data is not constant.
\begin{figure}[h]
\centering
\includegraphics[width=0.6\textwidth]{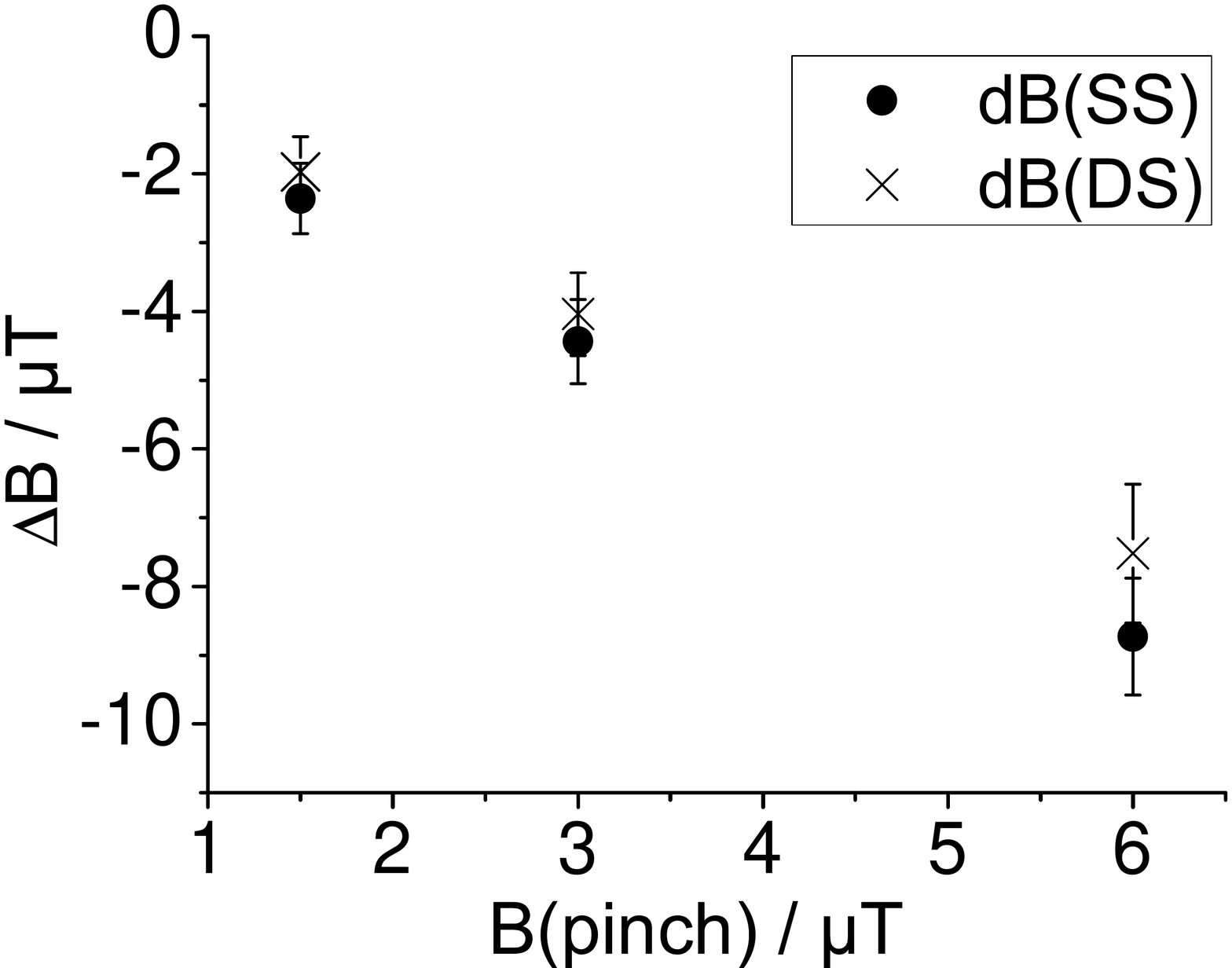}
\caption{\label{fig:dB} The difference $\Delta B$ between simulated magnetic field values and sample mean of the experimental values for the three magnetic field settings. Full dots: source sensor (SS), crosses: detector sensor (DS). The error bars shown refer to the standard deviation of each sample.}
\end{figure}
Taking the difference between the simulated values and the sample mean for the values given in Fig.\ref{fig:B-6T-SS-DS} a decrease can be observed (see  Fig.\ref{fig:dB})  which might be explained by magnetization of the construction steel used in the spectrometer hall.  

\newpage

\section{\label{sec:level1} Summary and Outlook} 

To control the magnetic field in the KATRIN main spectrometer area
a mobile magnetic field sensor unit is presented that moves along the LFCS inner belt and can take magnetic field samples. Such units can be installed at each LFCS ring and allows magnetic field sampling in areas that for safety reasons are only hardly or not at all accessible. The typical cycle time with 36 sampling stops is minutes.\par
The data presented in this paper might indicate magnetization effects in the vicinity of ring 17 caused by the detector solenoids. For a more detailed analysis of the magnetic field samples it is necessary to anchor the sensor coordinate systems into the global KATRIN coordinate system.
This can be achieved by incorporating inclination sensor data
and improved geometry data of the individual tracks into the analysis. Currently different types of inclinometers are studied.\par 
After multiple runs, the wheels of the MobS accumulate dust and slip can occur, which might lead to false position reading. Therefore we propose a combination of a toothed belt attached to the track and toothed wheels as a running gear of the MobS.\par
The magnetic field in the volume of the main spectrometer is free of rotation and therefore analytically connected with the field on the outside surface over the definition of a magnetic scalar Potential $V(x,y,z)$. It fulfills the Laplace equation which can be solved numerically by a finite difference method in which the magnetic surface samples are used as boundary values for computing the magnetic field everywhere inside the main spectrometer \cite{FLA, OSI1}. Hence it is advantageous to get large amounts of surface samples and the production of more MobS units that will run on further LFSC rings is under way. With these a more detailed determination of the magnetic field profile and the magnetization effects will be possible.  

\begin{table}[h]
\caption{\label{tab:sensors} Features of the flux gate sensors that are installed on the MobS. The sensors have also been calibrated by the company \cite{May}.}
\centering
\begin{tabular}{lcc}
\textrm{Sensor Type}&
\textrm{FLC3-70}&
\textrm{FL3-500}\\
\hline
Range & $\pm 500\mu $T & $\pm 1000\mu $ T \\
Accuracy & $\pm 1 \% \pm 0.5 \mu T$  & $\pm 0.5 \% $ \\
Resolution&1V / 35 $\mu T$ & 1V / 100 $\mu T $\\
Voltage Supply& 30 V & 30 V \\
Orthogonality &$< 1^{\circ}$& $< 0.5^{\circ}$
\end{tabular}

\end{table}


\vspace{0.5cm}

\noindent\textbf{\sffamily{AKNOWLEDGMENTS}}

\vspace{0.5cm}
The authors wish to express gratitude to the group for
Experimental Techniques of the Institute for Nuclear Physics
(IK) at KIT for highly efficient and competent support. Namely Prof. Dr. J. Blümer, Dr. F. Glück and Jan Reich for support and help.
Special thanks to the KATRIN detector group from University of Washington for supplying magnetic fields during the measurements.
Furthermore, we wish to thank Prof. Dr. E. W. Otten, Mainz University and Prof. Dr. Ch. Weinheimer, Münster University for helpful discussions and support.
In addition, we like to thank the University of Applied Sciences, Fulda and the Fachbereich Elektrotechnik und Informationstechnik, especially Prof. Dr. U. Rausch, for the enduring support for this work.

This work has been funded by the German Ministry
for Education and Research under the Project codes
05A11REA, 05A08RE1.

\end{document}